\documentclass[sigconf]{acmart}
\usepackage{multirow}

\AtBeginDocument{%
	\providecommand\BibTeX{{%
			\normalfont B\kern-0.5em{\scshape i\kern-0.25em b}\kern-0.8em\TeX}}}

\copyrightyear{2020}
\acmYear{2020}
\setcopyright{acmcopyright}\acmConference[SIGIR '20]{Proceedings of the 43rd International ACM SIGIR Conference on Research and Development in Information Retrieval}{July 25--30, 2020}{Virtual Event, China}
\acmBooktitle{Proceedings of the 43rd International ACM SIGIR Conference on Research and Development in Information Retrieval (SIGIR '20), July 25--30, 2020, Virtual Event, China}
\acmPrice{15.00}
\acmDOI{10.1145/3397271.3401269}
\acmISBN{978-1-4503-8016-4/20/07}

\begin{document}
\fancyhead{}

\title{Read what you need: Controllable Aspect-based Opinion Summarization of Tourist Reviews}

\author{Rajdeep Mukherjee}
\affiliation{\institution{IIT Kharagpur, India}}
\email{rajdeep1989@iitkgp.ac.in}

\author{Hari Chandana Peruri}
\affiliation{\institution{IIT Kharagpur, India}}
\email{chandupvsl@iitkgp.ac.in}

\author{Uppada Vishnu}
\affiliation{\institution{IIT Kharagpur, India}}
\email{vishnu14july@iitkgp.ac.in}

\author{Pawan Goyal}
\affiliation{\institution{IIT Kharagpur, India}}
\email{pawang@cse.iitkgp.ac.in}

\author{Sourangshu Bhattacharya}
\affiliation{\institution{IIT Kharagpur, India}}
\email{sourangshu@cse.iitkgp.ac.in}

\author{Niloy Ganguly}
\affiliation{\institution{IIT Kharagpur, India}}
\email{niloy@cse.iitkgp.ac.in}

\renewcommand{\shortauthors}{Rajdeep Mukherjee, et al.}

\begin{abstract}
	Manually extracting relevant aspects and opinions from large volumes of user-generated text is a time-consuming process. Summaries, on the other hand, help readers with limited time budgets to quickly consume the key ideas from the data. State-of-the-art approaches for multi-document summarization, however, do not consider user preferences while generating summaries. In this work, we argue the need and propose a solution for generating personalized aspect-based opinion summaries from large collections of online tourist reviews. We let our readers decide and control several attributes of the summary such as the length and specific aspects of interest among others. Specifically, we take an unsupervised approach to extract coherent aspects from tourist reviews posted on \textit{TripAdvisor}. We then propose an Integer Linear Programming (ILP) based extractive technique to select an informative subset of opinions around the identified aspects while respecting the user-specified values for various control parameters. Finally, we evaluate and compare our summaries using crowdsourcing and ROUGE-based metrics and obtain competitive results.
\end{abstract}




\maketitle

\section{introduction}\label{intro}
Tourism and Hospitality is an ever-growing industry and a crucial driver for the economic growth and development of any nation. Every year, millions of tourists travel across the world and share their pleasant/unpleasant experiences online through various platforms in the form of blogs and reviews. Future travelers and local service providers are benefited alike from this wide range of opinionated information for shaping their decisions \cite{MarreseTaylor2014AND}. However, manually going through all reviews and extracting relevant knowledge from them is an overwhelming task. Also, readers may have different levels of information needs which are not considered by the current state-of-the-art approaches for multi-document summarization \cite{Chu2019MeanSumAN, Nallapati2016SummaRuNNerAR}. We propose an \textit{unsupervised} and \textit{controllable} summarization framework which extracts aspect-based opinion summaries from huge corpora of tourist reviews, the shape and content of which can be customized according to reader's preferences.

\textit{Opinion Summarization}, as introduced in \cite{Hu2004MiningAS}, deals with (1) finding relevant aspects (such as \textit{amenities, culture}, etc.) about the entity being reviewed (here, \textit{place}) and identifying sentences which discuss them; (2) discovering sentiments expressed towards the identified aspects; and (3) generating a concise and digestible summary of opinions. While general-purpose summarization approaches \cite{Chu2019MeanSumAN, Nallapati2016SummaRuNNerAR} try to optimize the salience of the overall generated content, opinion summarization \cite{Condori2017OpinionSM} is more challenging as it is \textit{aspect-centric} and focuses on maximizing the diversity of opinions being covered in the final summary. \textit{Supervised} methods for opinion summarization such as \cite{Tian2019AspectAO} depend on large annotated datasets of document-summary pairs to train their models, which make them difficult to adapt across domains \cite{Brazinskas2019UnsupervisedMO}. Motivated by recent works such as \cite{Zhao2020WeaklySupervisedOS, Angelidis2018SummarizingOA}, we propose an \textit{unsupervised} extractive summarization framework built on top of an unsupervised aspect extraction module which makes our method generalizable to any domain.

Travel reviews posted online capture a wide range of human emotions owing to reviewers' social and cultural backgrounds. State-of-the-art methods for opinion summarization \cite{Brazinskas2019UnsupervisedMO, Condori2017OpinionSM} do not allow for personalization while generating summaries from such diverse range of opinionated text. We, however, argue that readers should be able to customize the shape and content of such summaries to suit their varying interests and time budgets. To our knowledge, only \cite{Amplayo2019InformativeAC, Fan2017ControllableAS} have motivated the need for controllable summarization. However, both of them are supervised techniques and require gold-standard summaries for training their models. Our proposed framework is unsupervised and extractive in nature which additionally lets our readers control several attributes of the summary such as its length and specific aspects of interest it must focus on. Motivated by \cite{Dash2019SummarizingUT}, our default summaries contain an equal proportion of opinions from male and female reviewers. We further make this ratio customizable to suit user preferences.

For our experiments, we create a dataset consisting of user reviews posted on \textit{TripAdvisor} (refer Section \ref{experiments}). First, we identify relevant aspects from the data following an unsupervised approach as proposed in \cite{He2017AnUN} (refer Section \ref{ate}). Then, we assign a score to each review sentence based on its salience. Finally, our ILP-based summarization framework extracts an optimal subset of sentences that best summarizes opinions about various aspects of the place being reviewed (refer Section \ref{summarization}). We achieve competitive performance on a variety of evaluation measures when compared against our baseline methods for unsupervised summarization (refer Section \ref{experiments}). We make our codes, datasets, survey instruments and a link to our web application publicly available at\\ \href{https://github.com/rajdeep345/ControllableSumm/}{https://github.com/rajdeep345/ControllableSumm/}.

\section{Related Work}\label{related}
We do not propose a new method for aspect extraction. However, as motivated in Section \ref{ate}, we need to limit the no. of aspect options, a control parameter in our proposed framework. Dependence of \textit{supervised} methods such as \cite{Xu2018DoubleEA} on considerable amount of labeled data and their inability to cluster the extracted aspect terms into coherent topics make them unsuitable for our purpose. We, therefore, follow an \textit{unsupervised} approach as proposed in \cite{He2017AnUN} to identify relevant aspects from our tourist-review corpus.

\textit{Supervised} methods for opinion summarization \cite{Condori2017OpinionSM} depend on significant volumes of training data  which is difficult to procure across domains \cite{Chu2019MeanSumAN, Brazinskas2019UnsupervisedMO}. Our ILP-based summarization framework, on the other hand, draws its motivation from \textit{unsupervised} approaches such as \cite{Zhao2020WeaklySupervisedOS, Angelidis2018SummarizingOA, Rudra2018IdentifyingSA}. None of these methods, however, facilitate readers to decide and control what they read. Though \cite{Amplayo2019InformativeAC, Fan2017ControllableAS} take user preferences into account, they are supervised techniques and require gold-standard summaries for training. Our proposed framework is unsupervised and extractive in nature. It further allows readers to control the shape and content of the generated summaries according to their preferences.

\section{Problem Formulation}\label{problem}
Let $ R^p = \{R_1^p, R_2^p, ..., R_n^p\} $ represent the set of $ n $ reviews available for a place $ p $, where each review $ R_i^p = \{R_{i1}^p, R_{i2}^p, ..., R_{im}^p\} $ represents a set of $ m $ sentences. $\ S^p = \cup_{i=1}^n\cup_{j=1}^m R_{ij}^p $ therefore represents the set of all review sentences available for the place. Let, the set of all aspects, as identified by our unsupervised aspect extraction module, be represented by $ A = \{A_1, A_2, A_3, ..., A_k\} $. In our proposed solution for controllable summarization, readers have the flexibility of selecting the aspects they are interested in. Let that set be represented by $ A^r\subseteq A $. Further, let $ L $ (words) represent the desired length of summary. The aim of our ILP-based summarization module is to select a subset $ s^p\subset S^p $ which best summarizes the reviews in $ R^p $ within $ L $ words while capturing the important opinions expressed towards the aspects in $ A^r $. As all users may not always be interested in reading fair summaries, hence the obtained summary should additionally take into account the desired ratio of comments from female and male reviewers as set by the reader.

\section{Aspect Identification}\label{ate} 
For the task of identifying relevant aspects, we follow an unsupervised \textit{Attention-based Aspect Extraction (ABAE)} technique as proposed in ~\cite{He2017AnUN}. ABAE is essentially an autoencoder-based topic model where the goal is to learn a set of $ K $ aspect embeddings without any supervision, where $ K $ represents the no. of topics/aspects to be identified. Given a set of review sentences as input, the model is trained to minimize the sentence reconstruction loss while learning to attend on the aspect words (please refer \cite{He2017AnUN} for details). Finally, each extracted topic is manually interpreted by looking at its representative words and assigned a genuine aspect label.

In order to quantify the interpretability of each identified topic cluster, we calculate its \textit{coherence score} as defined in ~\cite{He2017AnUN}. The value of $ K $, for which the average coherence score of all the clusters is maximum, is selected as the ideal no. of discoverable topics from the review corpus. However, based on the results of a survey with 30 participants (frequent travelers and interested in reading opinion summaries of tourist reviews), we find the ideal value of $ K $, obtained in our experiments, to be much higher than the ideal no. of aspects readers would like to choose from while generating the summaries. This establishes the need for clustering the identified \textit{fine-grained} topics into \textit{coarse-grained} aspect classes (Refer Section \ref{experiments} for results).
\section{Summarization Framework}\label{summarization}
\subsection{Opinion Scoring}\label{opin_scoring}
As a first step of opinion summarization, we assign a salience score to each review sentence based on the following three criteria:\\
\textbf{Readability Score:} In order to capture easily comprehensible opinions, we measure the readability of a sentence by means of its \textit{Flesch Reading Ease} score (0=very difficult and 100=very easy), which has been used as a widely-accepted metric since many years \cite{Yaneva2015EasyreadDA} for evaluating the \textit{simplicity of language}.\\
\textbf{Sentiment Strength:} We use the CoreNLP \textit{SentimentAnnotator} \cite{Manning2014TheSC} to obtain the sentiment polarity score of a sentence. It is an integer value in $ [0, 4] $ range, with $ 0 $ and $ 4 $ representing the most negative and positive sentiments, respectively. We obtain the sentiment strength as an absolute difference of this score with 2.\\
\textbf{Aspect-Relevance:} Each sentence $ s $ is assigned a relevance score based on how strongly it presents an opinion about a specific aspect. To account for the fact that a sentence might relate to more than one aspect, ~\cite{Zhao2020WeaklySupervisedOS} calculates this score as an average of its cosine similarity scores (at word level) with all the aspects in $ A $. We, however observe from the data that majority of the sentences (especially with strong sentiment polarity) focus on a particular aspect. Further, such averaging reduces the actual strength with which the sentence relates to this aspect. After preprocessing (mainly to discard irrelevant words such as pronouns and stop-words), we obtain the relevance of each word $ w_i $ with each of the aspects in $ A^r $ (by calculating cosine similarity between their embeddings) and finally take the maximum of these values to get the relevance score of the sentence, $ Relevance(s) $. Therefore, 
\begin{equation}
	\begin{aligned} \small
	Relevance(s) = \underset{1 \leq i \leq \vert s \vert}{\max} \; ( \; \underset{1 \leq j \leq \vert A^r \vert}{\max} \; cos(w_i, A_j) \; )
	\end{aligned}
\end{equation}
We multiply these three scores to obtain the opinion score of the sentence $ s $ denoted by $ Opin\_Score(s) $.

\subsection{Opinion Summarization}
The goal of our opinion summarization task is to maximize the collective salience of the selected subset of sentences while minimizing redundancy. Additionally, our framework allows readers to customize the summaries according to their needs. Let the desired length of summary be $ L $ words. Further, let $ fp $ represent the desired percentage of female opinions to be included in the summary. We formalize these objectives as a constrained maximization problem using ILP. In the following equations, $ l_i $ gives us the length of sentence $ s_i $ in words. $ f_i $ and $ m_i $ are complements of each other where $ f_i = 1 $ if $ s_i $ is an opinion made by a female reviewer and $ m_i = 1 $ if it is otherwise. $ x_i $ is a binary indicator variable whose value indicates whether to include sentence $ s_i $ in the final summary. $ sim_{ij} $ represents the cosine similarity between the embeddings of sentences $ s_i $ and $ s_j $. Similar to \cite{Zhao2020WeaklySupervisedOS}, equations \ref{Redundancy_Constraint1} and \ref{Redundancy_Constraint2} ensure that $ y_{ij} $, another binary variable, is $ 1 $ if and only if both $ x_i $ and $ x_j $ are $ 1 $. We use \textit{Gurobi} \cite{gurobi} to maximize the objective defined in Equation \ref{ILP}.
\begin{equation}\label{ILP}
\begin{aligned} \small
{\operatorname{argmax}} \; \sum_{i} Opin\_Score(s_i) \ x_i \ - \sum_{ij} sim_{ij} \ y_{ij} \ - C
\end{aligned}
\end{equation}
\begin{equation}\label{Fairness_Constraint}
\begin{aligned} \small
where \; \; C = \vert \; fp \sum_{i} m_i x_i - \; (1-fp) \; \sum_{i} f_i x_i \; \vert
\end{aligned}
\end{equation}
\begin{equation}\label{Length_Constraint}
\begin{aligned} \small
s.t. \; \; \sum_{i} l_i x_i \; \leq \; L
\end{aligned}
\end{equation}
\begin{equation}\label{Redundancy_Constraint1}
\begin{aligned} \small
\; \; y_{ij} \; \leq \; \frac{1}{2} (x_i + x_j ) \; \; \forall \ i,j
\end{aligned}
\end{equation}
\begin{equation}\label{Redundancy_Constraint2}
\begin{aligned} \small
\; \; y_{ij} \; \geq \; x_i + x_j -1 \; \; \forall \ i,j
\end{aligned}
\end{equation}
Equation \ref{Length_Constraint} restricts the length of the summary to $ L $ words. Apart from maximizing the collective opinion scores of the selected sentences, Equation \ref{ILP} tries to minimize their collective mutual similarity scores. It also minimizes the absolute difference between the desired ratio $ fp $ and the actual ratio of female to male comments in the final summary, as defined by Equation \ref{Fairness_Constraint}.

\section{Experimental Setup and Results}\label{experiments}
\noindent \textbf{Dataset:} We create a diverse dataset by collecting all (\textit{English}) reviews posted on \textit{TripAdvisor} (till July 30, 2019) for the New Seven Wonders of the World. For each review scraped, we consider the following fields for our analysis: \textit{id, text, user rating (1-5), \#likes (no. of likes received), username} and \textit{reviewer's location}. We use \textit{geopy} \footnote{\url{https://github.com/geopy/geopy}} to obtain reviewer's \textit{country of origin} from his/her location information. We further use the \textit{genderComputer} \footnote{\url{https://github.com/tue-mdse/genderComputer}} package to determine reviewer's \textit{gender} from his/her `username' and `country of origin'. As majority of usernames on TripAdvisor are not well-formed names, gender could be properly identified for around 20\% of the reviews, on average, across the seven places. Among these, we include the top 1000 liked reviews for each place in our final dataset, statistics of which are shown in Table \ref{tab:review_dataset_stats}.
\begin{table}
	\footnotesize
	\centering
	\caption{Tourist Review Dataset Statistics}
	\label{tab:review_dataset_stats}	
	\begin{tabular}{|c|l|l|}
		\hline
		\multirow{2}{*}{\textbf{Place}} & \multicolumn{2}{c|}{\textbf{Nos. of Reviews}} \\ \cline{2-3} 
		& \multicolumn{1}{c|}{\textbf{Female}} & \multicolumn{1}{c|}{\textbf{Male}} \\ \hline
		Whole data & 3164 (45.2\%) & 3836 (54.8\%) \\ \hline
		\multicolumn{1}{|l|}{The Roman Colosseum (Rome)} & 492 (49.2\%) & 508 (50.8\%) \\ \hline
		Christ the Redeemer (Brazil) & 445 (44.5\%) & 555 (55.5\%) \\ \hline
		Machu Picchu (Peru) & 456 (45.6\%) & 544 (54.4\%) \\ \hline
		Petra (Jordan) & 439 (43.9\%) & 561 (56.1\%) \\ \hline
		Taj Mahal (India) & 398 (39.8\%) & 602 (60.2\%) \\ \hline
		Chichen Itza (Mexico) & 482 (48.2\%) & 518 (51.8\%) \\ \hline
		Great Wall of China (China) & 452 (45.2\%) & 548 (54.8\%) \\ \hline
	\end{tabular}	
\end{table}

\noindent \textbf{Experiments on Aspect Identification: } We want aspect discovery to be an unsupervised one-time process. In order to obtain aspects/topics which are common across various tourist destinations, we merge the data for all the seven places under consideration and perform our experiments using default parameter settings as mentioned in \cite{He2017AnUN}. We experiment with several values of $ K $ ranging from 10 to 50 and obtain the best set of interpretable topics for $ K = 25 $. We manually label each of the identified topics with \textit{fine-grained} aspect labels. We finally categorize the obtained topics into \textit{coarse-grained} aspect classes and find them to be well-aligned with the \textit{tourism} literature \cite{MarreseTaylor2014AND}. Result of this many-to-many mapping is reported in Table \ref{tab:fga_category_mapping}. Since aspect extraction is not a major contribution of this work, we do not evaluate our obtained results against a gold-standard annotated corpus.
\tabcolsep=0.15cm
\begin{table} 
	\footnotesize
	\caption{Mapping between (coarse-grained) aspect classes and (fine-grained) inferred topics}	
	\label{tab:fga_category_mapping}
	\centering
	\begin{tabular}{|l|p{0.35\textwidth}|}
		\hline
		\textbf{Aspect Classes} & \multicolumn{1}{c|}{\textbf{Inferred Topics}} \\
		\hline
		Attractions & Architecture, Monuments, Surroundings, Events \\
		\hline
		Access & Travel infrastructure, Long-distance modes of transport, Entrance, Time of visiting\\
		\hline
		Activities & Photography, Events, Shopping\\
		\hline
		Amenities & Tour Guides, Hospitality, Vendors, Services, Information, Food, Accommodation \\
		\hline
		Culture & Hospitality, Events, Climate, History, Dress\\
		\hline
		Cost & Cost of visiting, Tickets\\
		\hline
		Negatives & Bad Experiences, Vendors\\
		\hline
		Miscellaneous & Nationality, Adjectives \\
		\hline
	\end{tabular}
\end{table}

\noindent \textbf{Experiments on Summarization:} While assigning salience scores to opinions in Section \ref{opin_scoring}, we obtain the aspect (class) embeddings as the average of \textit{Word2Vec} embeddings of the top 10 aspect terms from each topic belonging to that class. For calculating sentence similarity scores, we use Sentence-BERT~\cite{Reimers2019SentenceBERTSE}, a state-of-the-art method for obtaining sentence embeddings. 

We compare our results with two recent unsupervised and extractive summarization techniques, \textbf{FairSumm} \cite{Dash2019SummarizingUT} and \textbf{Centroid} \cite{Rossiello2017CentroidbasedTS};\footnote{We also considered \cite{Zhao2020WeaklySupervisedOS, Angelidis2018SummarizingOA} but their codes/files/instructions are yet to be released.} and with \textbf{Opinosis} \cite{Ganesan2010OpinosisAG}, an abstractive method for opinion summarization. As per standard practices, we employ ROUGE to evaluate the summaries. We note that the original reviews are themselves short textual accounts of user opinions about a place. Therefore, we rank them based on the \#likes received, and consider the text from the top 10 liked reviews for each place as a proxy for gold summaries. As none of our baselines facilitate aspect-based summary generation, we compare our \textit{default} summaries ($ L = 100 $, $ fp = 0.5 $, set of all aspects $ A $) and report ROUGE Precision scores in Table \ref{tab:rouge_comparison}, \textit{macro-averaged} across all places. We also conduct ablation studies to explore the effect of various constraints and opinion-scoring modules and report our results in Table \ref{tab:rouge_comparison}.

Since these scores are tentative due to unavailability of gold summaries, we perform a human evaluation of our summaries using crowdsourcing. Specifically, we create anonymous survey forms (details available in our \href{https://github.com/rajdeep345/ControllableSumm/}{\textit{Github}} repository) for each of the seven places. We divide each form into three sections to compare our general as well as aspect-specific summaries with those of the baselines. In each section, the four summaries are \textit{anonymized}
and appear in a completely \textit{randomized} order. We ask the responders to evaluate the summaries based on the following three questions: \textbf{Q1 (aspect-coverage)} \textit{Which of the summaries best captures the opinions about the specified aspects?} \textbf{Q2 (readability)} \textit{Which of the summaries is most readable?}, and \textbf{Q3 (diversity)} \textit{Which of the summaries contains least amount of repetitive information?} Overall, we received 183 responses to each question (close to 26 responses per place). Results, \textit{micro-averaged} across all seven places, are reported in Table \ref{tab:survey_results}.

\noindent \textbf{Results and Discussion:} As noted from Table \ref{tab:rouge_comparison}, though we perform better than \textit{Centroid} and \textit{FairSumm} on most of the ROUGE scores, we are considerably outperformed by \textit{Opinosis}. Upon eyeballing, we find out that \textit{Opinosis} summaries mainly contain short abstractive phrases which frequently occur in the reviews, thereby increasing the ROUGE scores. (Please visit our \href{https://github.com/rajdeep345/ControllableSumm/}{\textit{Github}} repository for a comparative analysis of different summaries). Our observations are further strengthened by the crowdsourcing-based evaluation results (Table \ref{tab:survey_results}) which clearly show that \textit{Opinosis} summaries perform poorly, more so on \textit{readability} and \textit{diversity} metrics. We further find from Table \ref{tab:survey_results}, that our summaries consistently outperform the baselines across all the three criteria. In Table \ref{tab:rouge_comparison}, we observe a drop in scores when comparing our \textit{default} summaries with those without the fairness constraint, highlighting the importance of maintaining fairness in summaries. We also compare our unconstrained (\textit{basic}) summaries with those obtained by removing one of the opinion-scoring metrics at a time. Further drop in scores establishes the importance of including \textit{readability score} and \textit{sentiment strength} in computing the salience of opinionated sentences.

\begin{table}[]
	\footnotesize
	\caption{Comparison of ROUGE Scores.}
	\label{tab:rouge_comparison}
	\centering
	\begin{tabular}{lccc}
		\hline
		\textbf{Methods} & \textbf{ROUGE-1} & \textbf{ROUGE-2} & \textbf{ROUGE-L} \\ \hline
		Opinosis & 74.8 & 21.5 & 49.6 \\
		Centroid & 62.3 & 14.2 & 38.8 \\
		FairSumm & \textbf{67.0} & 15.2 & 41.0 \\
		\hline
		\textbf{Our method} &  &  & \\
		\hline
		\quad with all constraints & 66.2 & 16.2 & \textbf{43.3} \\
		\quad w/o Fairness & 63.1 & 14.9 & 41.7 \\
		\quad w/o Redundancy & 65.8 & 16.2 & 43.2 \\
		\hline
		\quad w/o both constraints &  &  & \\
		\hline
		\qquad basic & 63.1 & 14.9 & 41.7 \\
		\qquad w/o Readability & 61.2 & \textbf{17.0} & 42.3 \\
		\qquad w/o Sentiment & 61.9 & 11.4 & 40.7 \\
		\qquad w/o both & 57.7 & 9.4 & 36.6 \\
		\hline
	\end{tabular}
\end{table}
\tabcolsep=0.15cm
\begin{table}[]
	\footnotesize
	\centering
	\caption{Results of the crowdsourcing-based evaluation, micro-averaged over all the seven places. Values indicate (\%) of times a method is preferred for a particular question.}
	\label{tab:survey_results}
	\begin{tabular}{|l|l|l|l|l|l|l|l|l|l|}
		\hline
		\multicolumn{1}{|c|}{\multirow{2}{*}{\textbf{Methods}}} & \multicolumn{3}{c|}{\textbf{All Aspects}} & \multicolumn{3}{c|}{\textbf{Access}} & \multicolumn{3}{c|}{\begin{tabular}[c]{@{}c@{}}\textbf{Amenities,} \\ \textbf{Culture}\end{tabular}} \\ \cline{2-10} 
		\multicolumn{1}{|c|}{} & \multicolumn{1}{c|}{\textbf{Q1}} & \multicolumn{1}{c|}{\textbf{Q2}} & \multicolumn{1}{c|}{\textbf{Q3}} & \multicolumn{1}{c|}{\textbf{Q1}} & \multicolumn{1}{c|}{\textbf{Q2}} & \multicolumn{1}{c|}{\textbf{Q3}} & \multicolumn{1}{c|}{\textbf{Q1}} & \multicolumn{1}{c|}{\textbf{Q2}} & \multicolumn{1}{c|}{\textbf{Q3}} \\ \hline
		Opinosis & 0.1 & 0.09 & 0.07 & 0.08 & 0.08 & 0.08 & 0.12 & 0.09 & 0.08 \\ \hline
		Centroid & 0.14 & 0.15 & 0.1 & 0.08 & 0.1 & 0.13 & 0.13 & 0.09 & 0.21 \\ \hline
		FairSumm & 0.13 & 0.15 & 0.18 & 0.23 & 0.28 & 0.23 & 0.13 & 0.26 & 0.15 \\ \hline
		\textbf{Our Method} & \textbf{0.63} & \textbf{0.61} & \textbf{0.65} & \textbf{0.61} & \textbf{0.54} & \textbf{0.56} & \textbf{0.62} & \textbf{0.56} & \textbf{0.56} \\ \hline
	\end{tabular}
\end{table}

\section{Conclusion and Future Work}\label{conclusion}
To our knowledge, this is the first attempt at producing personalized aspect-based opinion summaries, using an unsupervised extractive summarization framework, which additionally maintain fair representation of opinions from male and female reviewers (default setting). We motivate the need for such summaries and create a tourist review dataset for our experiments. We further establish the effectiveness of our framework by comparing it with recent unsupervised methods for opinion summarization. In future, we would like to experiment with the data of lesser known places.

\begin{acks}
	This research is supported by IMPRINT-2, a national initiative of the Ministry of Human Resource Development (MHRD), India.
\end{acks}

\bibliographystyle{ACM-Reference-Format}
\bibliography{main}

\end{document}